\documentclass[%
 preprint,
 amsmath,amssymb,
 aps, prmaterials
]{revtex4-2}

\usepackage{graphicx}
\usepackage{dcolumn}
\usepackage{bm}

\usepackage{color}

\begin{document}


\title{Valence-change-driven reduction of antiphase boundaries in spinel ferrite epitaxial films} 

\author{Kouki Takeo}
\affiliation{Department of Applied Physics, University of Tsukuba, 1-1-1, Tennodai, Tsukuba, Japan 305-8573}
\author{Eiji Kita}
\affiliation{Department of Applied Physics, University of Tsukuba, 1-1-1, Tennodai, Tsukuba, Japan 305-8573}
\author{Hideto Yanagihara}
\affiliation{Department of Applied Physics, University of Tsukuba, 1-1-1, Tennodai, Tsukuba, Japan 305-8573}
\affiliation{Tsukuba Research Center for
Energy Materials Science (TREMS), University of Tsukuba, 1-1-1 Tennodai,
Tsukuba, Ibaraki 305-8573, Japan}
\affiliation{Tsukuba Research Center for Organic-
Inorganic Quantum Spin Science and Technology (OIQSST), University of
Tsukuba, 1-1-1 Tennodai, Tsukuba, Ibaraki 305-8573, Japan}

\date{\today}

\begin{abstract} 
Antiphase boundaries (APBs) formed in thin films sometimes cause severe degradation of their physical properties. In particular, a high density of APBs in spinel ferrite films generates a non-negligible magnetic dead layer near the interface. 
In this study, we examined the effect of post-oxidation annealing in an oxygen plasma atmosphere on Co$_{0.125}$Fe$_{2.875}$O$_4$(001) thin films grown on MgO(001) as a model system. The thickness of the magnetic dead layer was found to be significantly reduced after post-oxidation, resulting in an increase in the saturation magnetization  and an improved squareness ratio. Dark-field transmission electron microscopy analysis revealed that the post-oxidation process increased the antiphase domain size, indicating a substantial reduction in APB density. Furthermore, reflection high-energy electron diffraction and x-ray diffraction measurements confirmed that the spinel crystal structure and epitaxial strain were preserved after post-oxidation.
These results suggest that post-oxidation proceeds through a topotactic solid-state reaction in which Fe$^{2.5+}$ ions are oxidized to Fe$^{3+}$, accompanied by cation rearrangement across APBs, thereby reducing APB density without degrading crystallinity and leading to improved magnetic properties in spinel ferrite epitaxial films.
\end{abstract}


\maketitle 
\section{Introduction}
Spinel ferrites exhibit a wide variety of magnetic and electrical properties and have attracted considerable interest for spintronic applications~\cite{Suzuki1996,Zheng2023,Wang2016,ElMasry2025,Ojima2018}. In particular, insulating spinel ferrites are important for spin-filter devices and spin-current transport phenomena because of their high Curie temperatures, ferrimagnetic ordering, and electrically insulating nature~\cite{Lders2006,Uchida2010,Venkateshvaran2009}. However, their magnetic properties are often severely degraded in thin-film form because of antiphase boundaries (APBs) formed during epitaxial growth on lattice-mismatched substrates~\cite{Eerenstein2002,Margulies1997}. At APBs, antiferromagnetic exchange coupling is frequently induced, resulting in local cancellation of magnetic moments and the formation of magnetic dead layers~\cite{Voogt1998,Li2021}.

MgO(001) substrates are widely used for spinel ferrite epitaxy because their doubled lattice constant (8.42~\AA{}) closely matches that of spinel ferrites. In cobalt ferrite ($a = 8.38$~\AA{}), the tensile strain induced by MgO promotes perpendicular magnetic anisotropy through magnetoelastic coupling~\cite{Niizeki2013,Tainosho2019,Onoda2021}. However, compared with spinel oxides, rocksalt-structured MgO lacks tetrahedral A-site cations and half of the octahedral B-site cations, although the oxygen sublattice remains nearly identical. As a result, spinel ferrites grown on MgO can maintain a continuous oxygen framework while allowing multiple registries of the cation sublattices, leading to the formation of APBs~\cite{Hibma1999}. In spinel ferrite thin films grown on MgO, APBs are known to exhibit several shift vectors, with the $1/4[110]$ shift being dominant~\cite{Celotto2003,McKenna2014}. In Co$_{0.75}$Fe$_{2.25}$O$_4$/MgO(001) thin films, APB-induced magnetic dead layers extending over several nanometers have been reported~\cite{Niizeki2013}.

Various approaches have been explored to suppress APBs in spinel ferrite thin films, including the use of lattice-matched substrates~\cite{Tainosho2019,Li2021,Regmi2022}, electric-field-assisted growth~\cite{Kumar2018}, and post-deposition annealing~\cite{Nonaka2023,Gilks2014,Eerenstein2003}. However, most previous studies have focused on Fe$_3$O$_4$, and strategies for reducing APBs in insulating ferrite thin films with strong perpendicular magnetic anisotropy remain limited.

In this study, we investigate the effect of post-oxidation annealing in an oxygen plasma atmosphere on Co$_{0.125}$Fe$_{2.875}$O$_4$(001) epitaxial films grown on MgO(001). We demonstrate that post-oxidation significantly reduces the APB density and magnetic dead layer thickness while preserving the spinel crystal structure and epitaxial strain. The results suggest that APB reduction proceeds through a topotactic oxidation process accompanied by cation rearrangement.

\section{Experimental Methods}
Cobalt ferrite (CFO) thin films were grown on cleaved MgO(001) substrates by reactive magnetron sputtering using a Co--Fe alloy target with a Co:Fe composition ratio of 1:23. To investigate the effect of post-oxidation, two types of as-grown films, referred to as sample~1 and sample~2, were prepared under different oxygen flow conditions during deposition. Sample~3 was subsequently obtained by post-oxidation annealing of sample~2 in an oxygen plasma atmosphere. The details of the samples are summarized in Table~\ref{tab:sample}.

\begin{table*}[hbt]
\caption{Growth and post-oxidation conditions, antiphase domain sizes, and film thicknesses for the CFO thin films investigated in this study.}
\label{tab:sample}
\begin{ruledtabular}
\begin{tabular}{cccccc}
Sample &
Description &
\shortstack[c]{Deposition\\O$_2$} &
\shortstack[c]{Annealing\\O$_2$} &
\shortstack[c]{Antiphase\\domain size} &
Thickness \\
&
&
(sccm) &
(sccm) &
(nm) &
(nm) \\
\hline
1 & As-grown & 4.0 & -- & $16.0 \pm 2.7$ & 20, 30 \\
2 & As-grown & 1.0 & -- & $18.3 \pm 2.4$ & 15, 20, 40, 60 \\
3 & Post-oxidized in O$_2$ plasma & 1.0 & 4.0 & $38.8 \pm 10$ & 15, 20, 40, 60 \\
\end{tabular}
\end{ruledtabular}
\end{table*}
During both deposition and annealing processes, the substrate temperature was maintained at 450~$^\circ$C, and the argon flow rate was fixed at 30~sccm. The chamber pressure was kept in the range of 0.5--0.6~Pa for all processes. Post-oxidation was carried out under the same ambient conditions as those used for sputter deposition, with the shutter between the substrate and the target closed. Multiple films with thicknesses ranging from 15 to 60~nm were prepared.

Structural characterization was carried out by \textit{in situ} reflection high-energy electron diffraction (RHEED) observations during film growth and X-ray diffraction (XRD) measurements  using Co~K$\alpha_1$ radiation ($\lambda = 1.789$~\AA) after deposition. The valence state of Fe was evaluated by conversion electron M\"ossbauer spectroscopy (CEMS) at room temperature. Transmission electron microscopy (TEM) was employed to observe the antiphase boundary (APB) domains. Magnetization curves were measured at room temperature using a vibrating sample magnetometer.

\section{Results and Discussion}
Electrical transport measurements performed immediately after deposition revealed that sample~1 and sample~3 exhibited insulating behavior, whereas sample~2 was electrically conductive. The resistivity of sample~2 was approximately $3\times10^{-2}$~$\Omega$\,cm.

\begin{figure}[ht]
    \centering
    \includegraphics[width=1\linewidth]{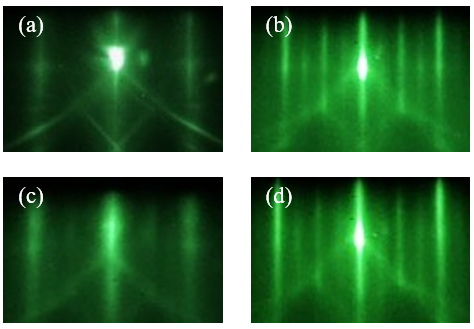}
    \caption{RHEED patterns observed with the electron beam incident along the [100] direction for (a) the MgO(001) substrate, (b) sample 1, (c) sample 2, and (d) sample 3.}
    \label{fig:RHEED}
\end{figure}

Figure \ref{fig:RHEED} shows the RHEED patterns taken along the MgO[100] azimuth for the MgO substrate and each CFO sample. Streak patterns were observed for all CFO films, indicating that all samples were epitaxially grown. Focusing on Figs. \ref{fig:RHEED}(b) and \ref{fig:RHEED}(d), clear streaks, Kikuchi lines, and surface reconstruction patterns were observed in both samples, suggesting high crystallinity and surface flatness \cite{ElMasry2025}.

\begin{figure}[hbt]
    \centering
    \includegraphics[width= .5\linewidth]{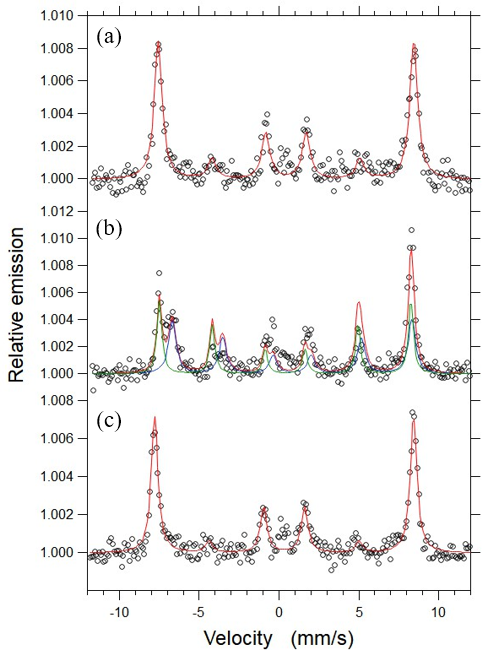}
    \caption{M\"ossbauer spectra measured at room temperature (RT). Open circles represent the experimental data, and solid lines indicate the fitting results.
(a) Sample 1: the red line represents the Fe$^{3+}$ component.
(b) Sample 2: the red line represents the total fit, the blue line the Fe$^{2.5+}$ component, and the green line the Fe$^{3+}$ component.
(c) Sample 3: the red line represents the Fe$^{3+}$ component.}
    \label{fig:CEMS}
\end{figure}

The valence states of iron in the samples at room temperature were evaluated by CEMS. The fitting parameters for each sample with a thickness of approximately 30 nm are summarized in Table~\ref{tab:CEMS}, and the corresponding M\"ossbauer spectra are shown in Fig.~\ref{fig:CEMS}.
All samples exhibited clear magnetic spectra without any contribution from $\alpha$-Fe. For sample~1 and sample~3, shown in Figs.~\ref{fig:CEMS}(a) and ~\ref{fig:CEMS}(c), respectively, six-line spectra characteristic of Fe$^{3+}$ were observed. From the intensity ratios of the spectra, the second and fifth peaks were nearly absent, indicating that the magnetic moments were oriented perpendicular to the film plane\cite{gutlich1971}. 

\begin{table}[hbt]
\caption{Summary of the fitting parameters obtained from the CEMS measurements. $H_{\mathrm{hf}}$ and IS denote the hyperfine field and isomer shift, respectively. For sample~2, the fitting was performed assuming two contributions originating from the tetrahedral A sites and octahedral B sites in the spinel structure. The labels A and B in parentheses indicate the values corresponding to each spectral component.}
\label{tab:CEMS}
\begin{ruledtabular}
\begin{tabular}{cccc}
Sample &
$\mu_0H_{\text{hf}}$ (T) &
$IS$ (mm/sec) &
Area (\%) \\
\hline
1 & 49.7 & 0.44 & 100 \\
2 & 48.9(A)/46.6(B) & 0.38(A)/0.83(B) & 52.1(A)/47.9(B) \\
3 & 50.4 & 0.32 & 100 \\
\end{tabular}
\end{ruledtabular}
\end{table}
In contrast, for sample~2 shown in Fig.~\ref{fig:CEMS}(b), an asymmetric spectrum composed of overlapping Fe$^{2.5+}$ and Fe$^{3+}$ magnetic components was observed. Fitting analysis revealed that sample~2 contains 52\% Fe$^{2.5+}$ and 48\% Fe$^{3+}$. Similar to magnetite, the presence of Fe$^{2.5+}$ is attributed to hopping conduction, which is consistent with the fact that sample~2 was the only electrically conductive sample. Furthermore, since the spectral intensity ratio was close to 3:2:1:1:2:3, the perpendicular component of magnetization in the film was found to be smaller than those in sample~1 and sample~3.It should also be noted that the linewidth of sample~3 was narrower than that of sample~1, suggesting improved local crystallinity in sample~3 after post-oxidation.

\begin{figure}[ht]
    \centering
    \includegraphics[width=1\linewidth]{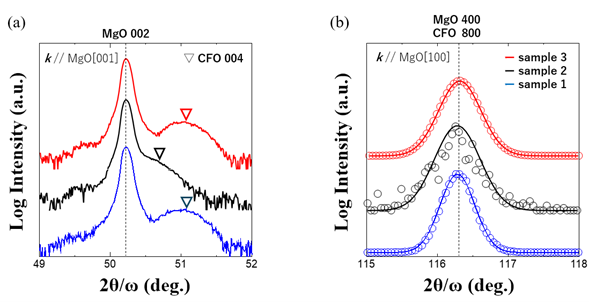}
    \caption{XRD patterns measured around the out-of-plane CFO(004) reflection (a) and the in-plane CFO(800) reflection (b). Solid curves represent the fitting results. For clarity, the data for samples 1, 2, and 3 are vertically shifted.}
    \label{fig:XRD}
\end{figure}

Figure~\ref{fig:XRD}(a) shows the XRD patterns measured along the MgO[001] direction for each sample. Since all observed diffraction peaks were indexed along the [001] direction, it was confirmed that all films were epitaxially grown on MgO(001) substrates, consistent with the RHEED results. In addition, the XRD patterns measured along the MgO[100] direction, shown in Fig.~\ref{fig:XRD}(b), revealed overlapping diffraction peaks from MgO and CFO, further indicating the epitaxial growth of the CFO films.
The out-of-plane lattice parameters were determined to be 8.28~\AA{} for sample~1 and sample~3, and 8.37~\AA{} for sample~2. These results indicate that the lattice parameters decreased after post-oxidation due to an increase in cation vacancies. Furthermore, sample~1 and sample~3 exhibited nearly identical lattice parameters. 
In contrast, along the [100] direction, the in-plane lattice parameters of the CFO films were constrained by the MgO substrate and therefore fixed at 8.42~\AA{}, corresponding to twice the lattice parameter of MgO. This value is larger than the out-of-plane lattice parameters measured along the [001] direction for all CFO samples prepared in this study, indicating that tensile strain was introduced by the substrate constraint.

The change in film thickness before and after post-oxidation was evaluated by X-ray reflectivity (XRR) measurements. The thicknesses of sample~2 and sample~3 were determined to be \(32.0 \pm 0.1\)~nm and \(33.7 \pm 0.7\)~nm, respectively, indicating that the film thickness increased by 5.4\% after post-oxidation.

Two possible factors can account for this thickness change. The first is the change in the unit-cell volume. In this case, the out-of-plane lattice parameter along the [001] direction decreased after post-oxidation. Specifically, the lattice parameter was reduced by 1.1\%.
The second factor is the increase in film volume associated with oxidation-induced changes in cation occupancy. As shown above, the CEMS results revealed that sample~2 contained 52\% Fe$^{2.5+}$, whereas in sample~3 all Fe$^{2.5+}$ ions were converted into Fe$^{3+}$. Assuming that the A sites are fully occupied by Fe$^{3+}$ and that the total numbers of Co and Fe ions remain unchanged during post-oxidation, the number of oxygen ions constituting the thin film is estimated to increase by approximately 9\% after oxidation\cite{Morishita2023}.
Based on this simple estimation, the film thickness is expected to increase by approximately 8\% after post-oxidation. However, the XRR measurements showed only a 5.4\% increase. This discrepancy is likely attributable to accumulated uncertainties arising from the XRR measurements, the fitting analysis of the M\"ossbauer spectra, and the simplified assumptions used in the calculations.

\begin{figure}[htb]
    \centering
    \includegraphics[width=1\linewidth]{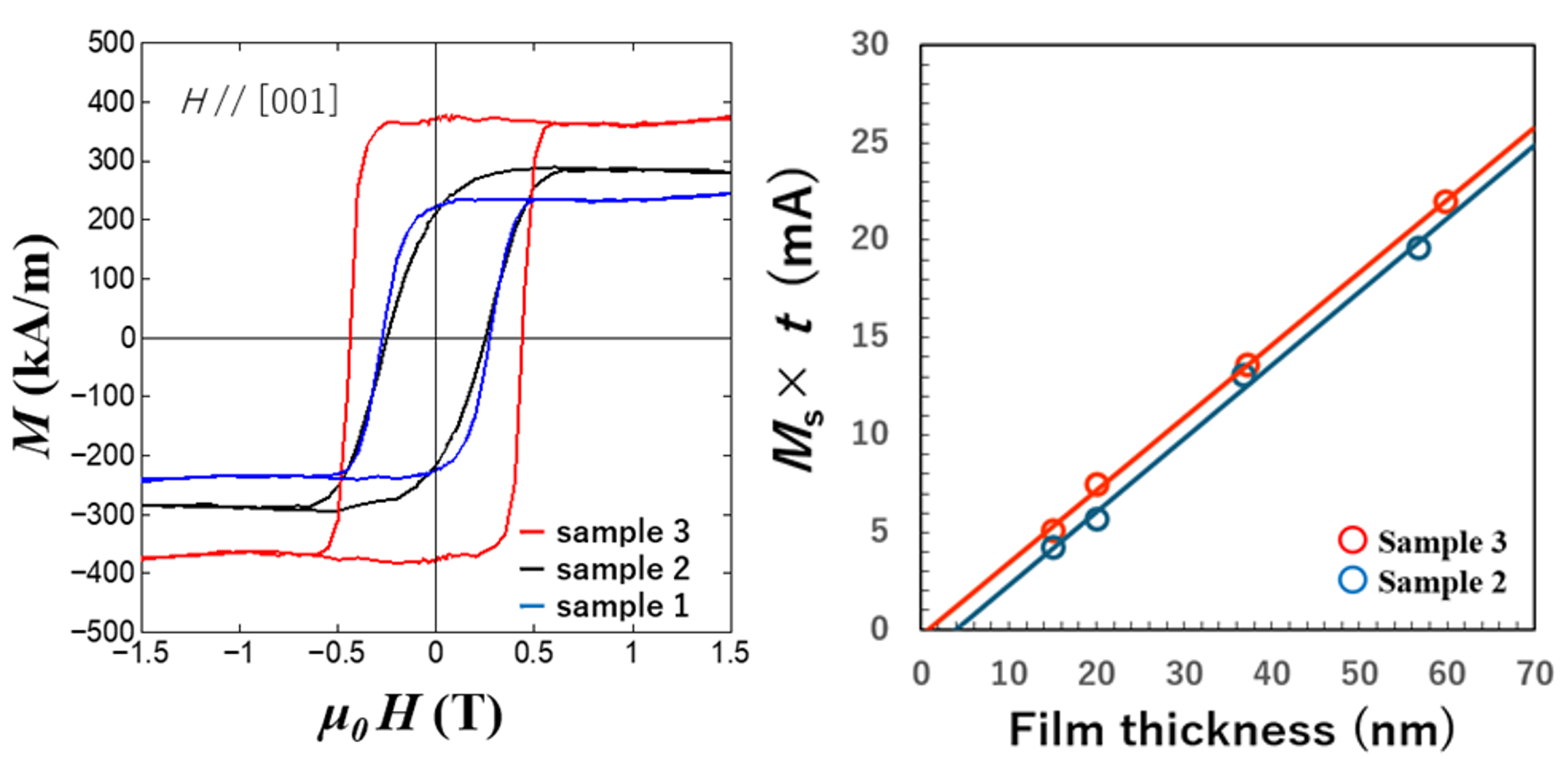}
    \caption{(a) Out-of-plane {\it MH} curves of the CFO films measured at room temperature (RT). (b) Film-thickness dependence of the saturation magnetization normalized by area. The {\it x}-intercept corresponds to the thickness of the magnetic dead layer.}
    \label{fig:MH}
\end{figure}

Next, we discuss the magnetic properties of the films. Figure~\ref{fig:MH}(a) shows the magnetization curves measured at room temperature for each CFO film with a thickness of approximately 20~nm. Focusing on sample~3, the saturation magnetization per unit volume, remanent magnetization, squareness ratio, and coercive field were all enhanced compared with those of sample~1 and sample~2. In particular, the saturation magnetization was significantly improved from 235~kA/m for sample~1 and 285~kA/m for sample~2 to 373~kA/m for sample~3, suggesting that the influence of the magnetic dead layer was reduced by the post-oxidation process. In addition, the squareness ratios of sample~1, sample~2, and sample~3 were 0.95, 0.74, and 0.99, respectively. These results are qualitatively consistent with the CEMS results, which suggest that the remanent magnetization in sample~1 and sample~3 is oriented more perpendicular to the film plane than that in sample~2. It should also be noted that sample~3 showed a relatively large coercive field of 0.44~T.
In order to estimate the thickness of the magnetic dead layer, two sets of sample~2 films with different thicknesses were prepared, and one film from each set was post-oxidized to produce sample~3. Figure~\ref{fig:MH}(b)  shows the thickness dependence of the saturation magnetization per unit area. By defining the intercept of the fitted linear line with the horizontal axis as the dead layer thickness, the dead layer thicknesses were estimated to be \(0.8 \pm 0.5\)~nm for sample~3 and \(3.9 \pm 1.1\)~nm for sample~2. The result for sample~2 is consistent with a previous report~\cite{Eerenstein2003}. 
In addition, the saturation magnetization estimated from the slope of the fitted line was 373~kA/m for sample~3 and 376~kA/m for sample~2, indicating that post-oxidation did not significantly change the intrinsic saturation magnetization.

\begin{figure}[hbt]
    \centering
    \includegraphics[width=1\linewidth]{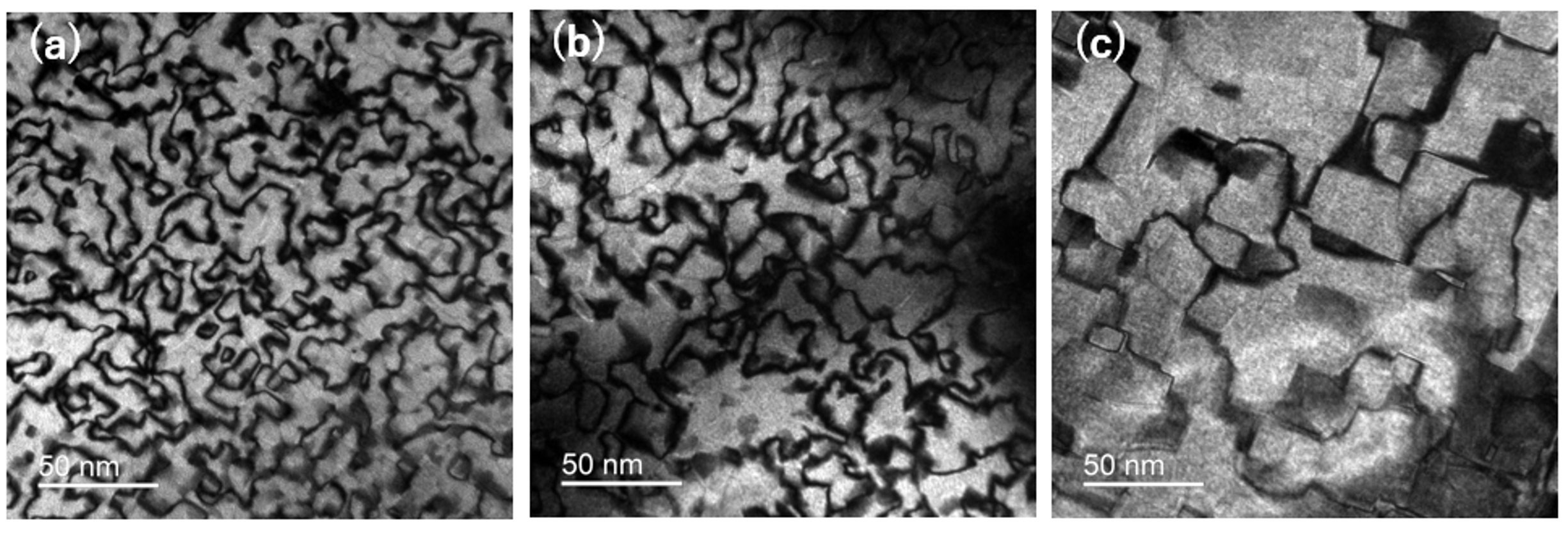}
    \caption{TEM dark-field images of 30-nm-thick CFO films obtained using the 220 diffraction spot for (a) sample 1, (b) sample 2, and (c) sample 3.}
    \label{fig:DFI}
\end{figure}

The APB density in each CFO film was evaluated by dark-field transmission electron microscopy (TEM) imaging. TEM specimens were prepared by dissolving the MgO substrates in a 4~wt\% ammonium sulfate aqueous solution at 70~$^\circ$C, followed by collecting the freestanding CFO thin films onto copper grids~\cite{Celotto2003}. Previous studies on Fe$_3$O$_4$/MgO(001) substrates reported that APBs appear along several types of boundary planes, with the \(1/4[110]\) shift being the dominant configuration~\cite{Celotto2003, McKenna2014}. Therefore, in this study, dark-field TEM images were obtained using the (220) diffraction spot. Figure~\ref{fig:DFI} shows the dark-field TEM images of each CFO thin film taken using the (220) reflection. In these images, bright and dark contrast can be observed, where the dark regions correspond to APBs. Sample~3  exhibits large antiphase domains and relatively straight domain boundaries. These features are similar to those reported in previous studies on annealed Fe$_3$O$_4$ films~\cite{Eerenstein2003}.
The antiphase domain sizes obtained by the intercept method are shown in Table~\ref{tab:sample}. The antiphase domain size of sample~3 was found to be more than twice as large as those of the other CFO samples, indicating that the use of conductive ferrite precursor films followed by post-oxidation is more effective in suppressing APBs than the direct growth of insulating ferrite films.
 
The enlargement of the antiphase domain size cannot be explained solely by thermally activated diffusion because sample~1, despite being grown at the same temperature, still exhibited a high APB density. These results suggest that the Fe valence change induced by post-oxidation plays an essential role in promoting APB reduction. During the post-oxidation process, Fe$^{2.5+}$ ions were oxidized to Fe$^{3+}$, as directly confirmed by the CEMS measurements. Such valence changes are expected to induce cation rearrangement while preserving the spinel crystal structure. Combined with the observed enlargement of the antiphase domains, these results suggest that APB reduction proceeds through a topotactic solid-state reaction accompanied by cation sublattice rearrangement. 

\section{Conclusions}

We demonstrated that post-oxidation treatment effectively suppresses antiphase boundaries (APBs) in insulating CFO epitaxial thin films. Post-oxidation was found to induce a topotactic oxidation reaction in which Fe$^{2.5+}$ ions were oxidized to Fe$^{3+}$ while preserving both the spinel crystal structure and the epitaxial strain. As a result, the perpendicular magnetic anisotropy originating from the epitaxial strain was maintained after post-oxidation, while the APB density was significantly reduced. The antiphase domain size increased by more than a factor of two, accompanied by a substantial reduction in the magnetic dead-layer thickness and a marked improvement in the magnetic properties.

The present results indicate that oxidation-induced cation rearrangement provides an effective pathway for APB annihilation while preserving the epitaxial framework and strain state of spinel ferrite films. Because this approach enables substantial defect reduction through a simple post-growth oxidation process without requiring higher growth temperatures, it offers a versatile strategy for defect engineering in spinel oxide epitaxial films. Such defect control is expected to facilitate the development of spinel-ferrite-based spintronic devices, including spin-filter devices~\cite{Lders2006}.

\begin{acknowledgments}
The TEM observations in this study were commissioned to the NIMS Open Facility. The authors thank Fumihiko Uesugi for his extensive support, ranging from image acquisition to the discussion.
M\"ossbauer studies were performed at Tandem Accelerator Complex, University of Tsukuba. This work was supported by the Japan Society for the Promotion of Science (JSPS) KAKENHI (22H04966, 23K26535, and 24H00408), and in part by the Advanced Research Infrastructure for Materials and Nanotechnology in Japan (ARIM) of the Ministry of Education, Culture, Sports, Science and Technology (MEXT), Japan (grant numbers JPMXP1224NM0187, JPMXP1225NM0209, JPMXP1224BA0008, JPMXP1225BA0030, and JPMXP1226BA0008).
\end{acknowledgments}



\bibliographystyle{apsrev4-2}
\bibliography{Takeo}

@article{Lders2006,
  title = {Spin filtering through ferrimagnetic NiFe2O4 tunnel barriers},
  volume = {88},
  ISSN = {1077-3118},
  url = {http://dx.doi.org/10.1063/1.2172647},
  DOI = {10.1063/1.2172647},
  number = {8},
  journal = {Applied Physics Letters},
  publisher = {AIP Publishing},
  author = {L\"{u}ders,  U. and Bibes,  M. and Bouzehouane,  K. and Jacquet,  E. and Contour,  J.-P. and Fusil,  S. and Bobo,  J.-F. and Fontcuberta,  J. and Barthélémy,  A. and Fert,  A.},
  year = {2006},
  month = feb 
}

@article{Uchida2010,
  title = {Spin Seebeck insulator},
  volume = {9},
  ISSN = {1476-4660},
  url = {http://dx.doi.org/10.1038/nmat2856},
  DOI = {10.1038/nmat2856},
  number = {11},
  journal = {Nature Materials},
  publisher = {Springer Science and Business Media LLC},
  author = {Uchida,  K. and Xiao,  J. and Adachi,  H. and Ohe,  J. and Takahashi,  S. and Ieda,  J. and Ota,  T. and Kajiwara,  Y. and Umezawa,  H. and Kawai,  H. and Bauer,  G. E. W. and Maekawa,  S. and Saitoh,  E.},
  year = {2010},
  month = sep,
  pages = {894–897}
}

@article{Venkateshvaran2009,
  title = {Epitaxial<mml:math xmlns:mml="http://www.w3.org/1998/Math/MathML" display="inline"><mml:mrow><mml:msub><mml:mrow><mml:mtext>Zn</mml:mtext></mml:mrow><mml:mi>x</mml:mi></mml:msub><mml:msub><mml:mrow><mml:mtext>Fe</mml:mtext></mml:mrow><mml:mrow><mml:mn>3</mml:mn><mml:mo>−</mml:mo><mml:mi>x</mml:mi></mml:mrow></mml:msub><mml:msub><mml:mtext>O</mml:mtext><mml:mn>4</mml:mn></mml:msub></mml:mrow></mml:math>thin films: A spintronic material with tunable electrical and magnetic properties},
  volume = {79},
  ISSN = {1550-235X},
  url = {http://dx.doi.org/10.1103/PhysRevB.79.134405},
  DOI = {10.1103/physrevb.79.134405},
  number = {13},
  journal = {Physical Review B},
  publisher = {American Physical Society (APS)},
  author = {Venkateshvaran,  Deepak and Althammer,  Matthias and Nielsen,  Andrea and Gepr\"{a}gs,  Stephan and Ramachandra Rao,  M. S. and Goennenwein,  Sebastian T. B. and Opel,  Matthias and Gross,  Rudolf},
  year = {2009},
  month = apr 
}

@article{Suzuki1996,
  title = {Structure and magnetic properties of epitaxial spinel ferrite thin films},
  volume = {68},
  ISSN = {1077-3118},
  url = {http://dx.doi.org/10.1063/1.116601},
  DOI = {10.1063/1.116601},
  number = {5},
  journal = {Applied Physics Letters},
  publisher = {AIP Publishing},
  author = {Suzuki,  Y. and van Dover,  R. B. and Gyorgy,  E. M. and Phillips,  Julia M. and Korenivski,  V. and Werder,  D. J. and Chen,  C. H. and Cava,  R. J. and Krajewski,  J. J. and Peck,  W. F. and Do,  K. B.},
  year = {1996},
  month = jan,
  pages = {714–716}
}

@article{Zheng2023,
  title = {Ultra-thin lithium aluminate spinel ferrite films with perpendicular magnetic anisotropy and low damping},
  volume = {14},
  ISSN = {2041-1723},
  url = {http://dx.doi.org/10.1038/s41467-023-40733-9},
  DOI = {10.1038/s41467-023-40733-9},
  number = {1},
  journal = {Nature Communications},
  publisher = {Springer Science and Business Media LLC},
  author = {Zheng,  Xin Yu and Channa,  Sanyum and Riddiford,  Lauren J. and Wisser,  Jacob J. and Mahalingam,  Krishnamurthy and Bowers,  Cynthia T. and McConney,  Michael E. and N’Diaye,  Alpha T. and Vailionis,  Arturas and Cogulu,  Egecan and Ren,  Haowen and Galazka,  Zbigniew and Kent,  Andrew D. and Suzuki,  Yuri},
  year = {2023},
  month = aug 
}

@article{Wang2016,
  title = {Lightweight NiFe2O4 with controllable 3D network structure and enhanced microwave absorbing properties},
  volume = {6},
  ISSN = {2045-2322},
  url = {http://dx.doi.org/10.1038/srep37892},
  DOI = {10.1038/srep37892},
  number = {1},
  journal = {Scientific Reports},
  publisher = {Springer Science and Business Media LLC},
  author = {Wang,  Fen and Wang,  Xing and Zhu,  Jianfeng and Yang,  Haibo and Kong,  Xingang and Liu,  Xiao},
  year = {2016},
  month = nov 
}

@article{ElMasry2025,
  title = {Cobalt,  nickel and zinc spinel ferrites with high transmittance and UV-blocking for advanced optical applications},
  volume = {15},
  ISSN = {2045-2322},
  url = {http://dx.doi.org/10.1038/s41598-025-99604-6},
  DOI = {10.1038/s41598-025-99604-6},
  number = {1},
  journal = {Scientific Reports},
  publisher = {Springer Science and Business Media LLC},
  author = {El-Masry,  Mai M. and Arman,  M. M.},
  year = {2025},
  month = may 
}

@article{Ojima2018,
  title = {RHEED oscillations in spinel ferrite epitaxial films grown by conventional planar magnetron sputtering},
  volume = {8},
  ISSN = {2158-3226},
  url = {http://dx.doi.org/10.1063/1.5012133},
  DOI = {10.1063/1.5012133},
  number = {4},
  journal = {AIP Advances},
  publisher = {AIP Publishing},
  author = {Ojima,  T. and Tainosho,  T. and Sharmin,  S. and Yanagihara,  H.},
  year = {2018},
  month = apr 
}

@article{Eerenstein2002,
  title = {Origin of the increased resistivity in epitaxial<mml:math xmlns:mml="http://www.w3.org/1998/Math/MathML" display="inline"><mml:mrow><mml:msub><mml:mrow><mml:mi mathvariant="normal">Fe</mml:mi></mml:mrow><mml:mrow><mml:mn>3</mml:mn></mml:mrow></mml:msub></mml:mrow><mml:mrow><mml:msub><mml:mrow><mml:mi mathvariant="normal">O</mml:mi></mml:mrow><mml:mrow><mml:mn>4</mml:mn></mml:mrow></mml:msub></mml:mrow></mml:math>films},
  volume = {66},
  ISSN = {1095-3795},
  url = {http://dx.doi.org/10.1103/PhysRevB.66.201101},
  DOI = {10.1103/physrevb.66.201101},
  number = {20},
  journal = {Physical Review B},
  publisher = {American Physical Society (APS)},
  author = {Eerenstein,  W. and Palstra,  T. T. M. and Hibma,  T. and Celotto,  S.},
  year = {2002},
  month = nov 
}

@article{Margulies1997,
  title = {Origin of the Anomalous Magnetic Behavior in Single Crystal<mml:math xmlns:mml="http://www.w3.org/1998/Math/MathML" display="inline"><mml:mrow><mml:msub><mml:mrow><mml:mi>Fe</mml:mi></mml:mrow><mml:mrow><mml:mn>3</mml:mn></mml:mrow></mml:msub></mml:mrow><mml:mrow><mml:msub><mml:mrow><mml:mi>O</mml:mi></mml:mrow><mml:mrow><mml:mn>4</mml:mn></mml:mrow></mml:msub></mml:mrow></mml:math>Films},
  volume = {79},
  ISSN = {1079-7114},
  url = {http://dx.doi.org/10.1103/PhysRevLett.79.5162},
  DOI = {10.1103/physrevlett.79.5162},
  number = {25},
  journal = {Physical Review Letters},
  publisher = {American Physical Society (APS)},
  author = {Margulies,  D. T. and Parker,  F. T. and Rudee,  M. L. and Spada,  F. E. and Chapman,  J. N. and Aitchison,  P. R. and Berkowitz,  A. E.},
  year = {1997},
  month = dec,
  pages = {5162–5165}
}

@article{Voogt1998,
  title = {Superparamagnetic behavior of structural domains in epitaxial ultrathin magnetite films},
  volume = {57},
  ISSN = {1095-3795},
  url = {http://dx.doi.org/10.1103/PhysRevB.57.R8107},
  DOI = {10.1103/physrevb.57.r8107},
  number = {14},
  journal = {Physical Review B},
  publisher = {American Physical Society (APS)},
  author = {Voogt,  F. C. and Palstra,  T. T. M. and Niesen,  L. and Rogojanu,  O. C. and James,  M. A. and Hibma,  T.},
  year = {1998},
  month = apr,
  pages = {R8107–R8110}
}

@article{Li2021,
  title = {Atomic Structure and Electron Magnetic Circular Dichroism of Individual Rock Salt Structure Antiphase Boundaries in Spinel Ferrites},
  volume = {31},
  ISSN = {1616-3028},
  url = {http://dx.doi.org/10.1002/adfm.202008306},
  DOI = {10.1002/adfm.202008306},
  number = {21},
  journal = {Advanced Functional Materials},
  publisher = {Wiley},
  author = {Li,  Zhuo and Lu,  Jinlian and Jin,  Lei and Rusz,  Ján and Kocevski,  Vancho and Yanagihara,  Hideto and Kita,  Eiji and Mayer,  Joachim and Dunin‐Borkowski,  Rafal E. and Xiang,  Hongjun and Zhong,  Xiaoyan},
  year = {2021},
  month = mar 
}

@article{Niizeki2013,
    author = {Niizeki, Tomohiko and Utsumi, Yuji and Aoyama, Ryohei and Yanagihara, Hideto and Inoue, Jun-ichiro and Yamasaki, Yuichi and Nakao, Hironori and Koike, Kazuyuki and Kita, Eiji},
    title = {Extraordinarily large perpendicular magnetic anisotropy in epitaxially strained cobalt-ferrite CoxFe3−xO4(001) (x = 0.75, 1.0) thin films},
    journal = {Applied Physics Letters},
    volume = {103},
    number = {16},
    pages = {162407},
    year = {2013},
    month = {10},
    issn = {0003-6951},
    doi = {10.1063/1.4824761},
}

@article{Onoda2021,
  title = {Strain Engineering of Magnetic Anisotropy in Epitaxial Films of Cobalt Ferrite},
  volume = {8},
  ISSN = {2196-7350},
  url = {http://dx.doi.org/10.1002/admi.202101034},
  DOI = {10.1002/admi.202101034},
  number = {23},
  journal = {Advanced Materials Interfaces},
  publisher = {Wiley},
  author = {Onoda,  Hiroshige and Sukegawa,  Hiroaki and Inoue,  Jun‐Ichiro and Yanagihara,  Hideto},
  year = {2021},
  month = nov 
}

@article{Hibma1999,
    author = {Hibma, T. and Voogt, F. C. and Niesen, L. and van der Heijden, P. A. A. and de Jonge, W. J. M. and Donkers, J. J. T. M. and van der Zaag, P. J.},
    title = {Anti-phase domains and magnetism in epitaxial magnetite layers},
    journal = {Journal of Applied Physics},
    volume = {85},
    number = {8},
    pages = {5291-5293},
    year = {1999},
    month = {04},
    issn = {0021-8979},
    doi = {10.1063/1.369857},
    url = {https://doi.org/10.1063/1.369857},
}

@article{Celotto2003,
  title = {Characterization of anti-phase boundaries in epitaxial magnetite films},
  volume = {36},
  ISSN = {1434-6036},
  url = {http://dx.doi.org/10.1140/epjb/e2003-00344-7},
  DOI = {10.1140/epjb/e2003-00344-7},
  number = {2},
  journal = {The European Physical Journal B - Condensed Matter},
  publisher = {Springer Science and Business Media LLC},
  author = {Celotto,  S. and Eerenstein,  W. and Hibma,  T.},
  year = {2003},
  month = nov,
  pages = {271–279}
}

@article{McKenna2014,
  title = {Atomic-scale structure and properties of highly stable antiphase boundary defects in Fe3O4},
  volume = {5},
  ISSN = {2041-1723},
  url = {http://dx.doi.org/10.1038/ncomms6740},
  DOI = {10.1038/ncomms6740},
  number = {1},
  journal = {Nature Communications},
  publisher = {Springer Science and Business Media LLC},
  author = {McKenna,  Keith P. and Hofer,  Florian and Gilks,  Daniel and Lazarov,  Vlado K. and Chen,  Chunlin and Wang,  Zhongchang and Ikuhara,  Yuichi},
  year = {2014},
  month = dec 
}

@article{Regmi2022,
    author = {Regmi, Sudhir and Li, Zhong and KC, Shambhu and Mahat, Rabin and Rastogi, Ankur and Datta, Ranjan and Gupta, Arunava},
    title = {Structural and magnetic properties of CoFe2O4 thin films grown on isostructural lattice-matched substrates},
    journal = {Applied Physics Letters},
    volume = {121},
    number = {10},
    pages = {102401},
    year = {2022},
    month = {09},
    issn = {0003-6951},
    doi = {10.1063/5.0107242},
    url = {https://doi.org/10.1063/5.0107242},
}

@article{Kumar2018,
  title = {Effect of 
in situ
 electric-field-assisted growth on antiphase boundaries in epitaxial 
<mml:math xmlns:mml="http://www.w3.org/1998/Math/MathML"><mml:mrow><mml:msub><mml:mi mathvariant="normal">Fe</mml:mi><mml:mn>3</mml:mn></mml:msub><mml:msub><mml:mi mathvariant="normal">O</mml:mi><mml:mn>4</mml:mn></mml:msub></mml:mrow></mml:math>
 thin films on MgO},
  volume = {2},
  ISSN = {2475-9953},
  url = {http://dx.doi.org/10.1103/PhysRevMaterials.2.054407},
  DOI = {10.1103/physrevmaterials.2.054407},
  number = {5},
  journal = {Physical Review Materials},
  publisher = {American Physical Society (APS)},
  author = {Kumar,  Ankit and Wetterskog,  Erik and Lewin,  Erik and Tai,  Cheuk-Wai and Akansel,  Serkan and Husain,  Sajid and Edvinsson,  Tomas and Brucas,  Rimantas and Chaudhary,  Sujeet and Svedlindh,  Peter},
  year = {2018},
  month = may 
}

@article{Nonaka2023,
  title = {Origin of magnetically dead layers in spinel ferrites 
<mml:math xmlns:mml="http://www.w3.org/1998/Math/MathML"><mml:mrow><mml:mi>M</mml:mi><mml:msub><mml:mi>Fe</mml:mi><mml:mn>2</mml:mn></mml:msub><mml:msub><mml:mi mathvariant="normal">O</mml:mi><mml:mn>4</mml:mn></mml:msub></mml:mrow></mml:math>
 grown on 
<mml:math xmlns:mml="http://www.w3.org/1998/Math/MathML"><mml:mrow><mml:msub><mml:mi>Al</mml:mi><mml:mn>2</mml:mn></mml:msub><mml:msub><mml:mi mathvariant="normal">O</mml:mi><mml:mn>3</mml:mn></mml:msub></mml:mrow></mml:math>
: Effects of postdeposition annealing studied by XMCD},
  volume = {7},
  ISSN = {2475-9953},
  url = {http://dx.doi.org/10.1103/PhysRevMaterials.7.044413},
  DOI = {10.1103/physrevmaterials.7.044413},
  number = {4},
  journal = {Physical Review Materials},
  publisher = {American Physical Society (APS)},
  author = {Nonaka,  Yosuke and Wakabayashi,  Yuki K. and Shibata,  Goro and Sakamoto,  Shoya and Ikeda,  Keisuke and Chi,  Zhendong and Wan,  Yuxuan and Suzuki,  Masahiro and Tanaka,  Arata and Tanaka,  Masaaki and Fujimori,  Atsushi},
  year = {2023},
  month = apr 
}

@article{Gilks2014,
    author = {Gilks, D. and Lari, L. and Matsuzaki, K. and Hosono, H. and Susaki, T. and Lazarov, V. K.},
    title = {Structural study of Fe3O4(111) thin films with bulk like magnetic and magnetotransport behaviour},
    journal = {Journal of Applied Physics},
    volume = {115},
    number = {17},
    pages = {17C107},
    year = {2014},
    month = {01},
    issn = {0021-8979},
    doi = {10.1063/1.4862524},
    url = {https://doi.org/10.1063/1.4862524},
}

@article{Eerenstein2003,
  title = {Diffusive motion of antiphase domain boundaries in<mml:math xmlns:mml="http://www.w3.org/1998/Math/MathML" display="inline"><mml:mrow><mml:msub><mml:mrow><mml:mi mathvariant="normal">Fe</mml:mi></mml:mrow><mml:mrow><mml:mn>3</mml:mn></mml:mrow></mml:msub></mml:mrow><mml:mrow><mml:msub><mml:mrow><mml:mi mathvariant="normal">O</mml:mi></mml:mrow><mml:mrow><mml:mn>4</mml:mn></mml:mrow></mml:msub></mml:mrow></mml:math>films},
  volume = {68},
  ISSN = {1095-3795},
  url = {http://dx.doi.org/10.1103/PhysRevB.68.014428},
  DOI = {10.1103/physrevb.68.014428},
  number = {1},
  journal = {Physical Review B},
  publisher = {American Physical Society (APS)},
  author = {Eerenstein,  W. and Palstra,  T. T. M. and Hibma,  T. and Celotto,  S.},
  year = {2003},
  month = jul 
}

@article{Morishita2023,
  title = {Control of conductivity in Fe-rich cobalt-ferrite thin films with perpendicular magnetic anisotropy},
  volume = {7},
  ISSN = {2475-9953},
  url = {http://dx.doi.org/10.1103/PhysRevMaterials.7.054402},
  DOI = {10.1103/physrevmaterials.7.054402},
  number = {5},
  journal = {Physical Review Materials},
  publisher = {American Physical Society (APS)},
  author = {Morishita,  Masaya and Ichikawa,  Tomoyuki and Tanaka,  Masaaki A. and Furuta,  Motoharu and Mashimo,  Daisuke and Honda,  Syuta and Okabayashi,  Jun and Mibu,  Ko},
  year = {2023},
  month = may 
}

@article{Tainosho2019,
    author = {Tainosho, Takeshi and Inoue, Jun-ichiro and Sharmin, Sonia and Takeguchi, Masaki and Kita, Eiji and Yanagihara, Hideto},
    title = {Large negative uniaxial magnetic anisotropy in highly distorted Co-ferrite thin films},
    journal = {Applied Physics Letters},
    volume = {114},
    number = {9},
    pages = {092408},
    year = {2019},
    month = {03},
    issn = {0003-6951},
    doi = {10.1063/1.5064845},
    url = {https://doi.org/10.1063/1.5064845},
}

@book{gutlich1971,
  title={M\"ossbauer Spectroscopy and Transition Metal Chemistry},
  author={G\"utlich, P and Bill E and Trautwein, A X},
  number={},
  year={2011},
  publisher={Springer, Heidelberg}}

\end{document}